\documentclass[
10pt,
showpacs,preprintnumbers,nofootinbib,
 amsmath,amssymb,
 aps,
prl,twocolumn,groupedaddress,superscriptaddress,
showkeys
]{revtex4-1}
\usepackage{graphicx}
\usepackage{dcolumn}
\usepackage{bm}
\usepackage[colorlinks=true,urlcolor=blue,citecolor=blue,breaklinks=true]{hyperref}
\usepackage{epsfig}

\begin{document}


\title{How many-body correlations and $\alpha$-clustering shape $^6$He}

\author{Carolina Romero-Redondo}
 \email{romeroredond1@llnl.gov}
 \affiliation{Lawrence Livermore National Laboratory, P.O. Box 808, L-414, Livermore, California 94551, USA}
\author{Sofia Quaglioni}
 \email{quaglioni1@llnl.gov}
 \affiliation{Lawrence Livermore National Laboratory, P.O. Box 808, L-414, Livermore, California 94551, USA}
 \author{Petr Navr\'atil}
 \email{navratil@triumf.ca}
 \affiliation{TRIUMF, 4004 Wesbrook Mall, Vancouver, British Columbia, V6T 2A3, Canada}
 \author{Guillaume Hupin}
 \affiliation{CEA, DAM, DIF, F-91297 Arpajon, France.}

\date{\today}

\begin{abstract}
The Borromean $^6$He nucleus is 
an exotic system characterized by two ÔhaloÕ neutrons orbiting around a compact $^4$He (or $\alpha$) core, in which the binary subsystems are unbound. The simultaneous reproduction of its small binding energy and extended matter and point-proton radii has been a challenge for  {\em ab initio} theoretical calculations  based on traditional bound-state methods. Using soft nucleon-nucleon interactions based on chiral effective field theory potentials, we show that supplementing the model space with $^4$He+$n$+$n$ cluster degrees of freedom largely solves this issue.   
We analyze the role played by the $\alpha$-clustering and many-body correlations, and study the dependence of the energy spectrum on the resolution scale of the interaction.
\end{abstract}
\pacs{21.60.De, 25.10.+s, 27.20.+n}
\maketitle

%
\paragraph{Introduction.}
Achieving a comprehensive and unified treatment of many-body correlations and clustering in atomic nuclei constitutes a frontier for contemporary nuclear theory. 
A light exotic nucleus that has been challenging our understanding of such complex phenomena based on 
nucleonic degrees of freedom
and high-quality models of their interactions (i.e., within an {\em ab initio} framework) is Helium-6 ($^6$He). 
This is a 
prominent example of  
Borromean quantum `halo', i.e.\ a weakly-bound state of three particles 
($\alpha$+$n$+$n$) otherwise unbound in pairs, characterized by ``large probability of configurations within classically forbidden regions of space" \cite{RevModPhys.76.215}.  
In the last few years, its binding energy~\cite{PhysRevLett.108.052504} and charge radius~\cite{PhysRevLett.93.142501} have been experimentally determined with high precision. 
The $^6$He ground state (g.s.)\ is also of great interest for tests of fundamental interactions and symmetries. 
Precision measurements of its $\beta$-decay half life have recently 
taken place~\cite{PhysRevLett.108.122502} and efforts are under way to determine the angular 
correlation between the emitted electron and neutrino \cite{garcia_private}.    
To date, traditional {\em ab initio} bound-state calculations 
can successfully describe  the interior of the $^6$He wave function
\cite{Pieper2008,Bacca:2012up,Saaf2014,Caprio2014,Constantinou2016}, 
but are unable to fully account for its three-cluster 
asymptotic behavior. 
At the same time, the only {\em ab initio} study 
of $\alpha$+$n$+$n$
dynamics naturally explains the asymptotic configurations, but underbinds the $^6$He g.s.\ owing to missing 
many-body correlations \cite{Quaglioni2013,Romero-Redondo:2014fya}.
As a result, a comprehensive description of the $^6$He g.s.\ properties is still missing.  
   
In this Letter we present a study of the $^6$He g.s.\ in which both six-body correlations and clustering are 
successfully addressed by means of 
the no-core shell model with continuum (NCSMC)~\cite{1402-4896-91-5-053002}. 
This approach, 
introduced to describe binary processes starting from two-~\cite{PhysRevLett.110.022505,PhysRevC.87.034326} 
and later three-body~\cite{PhysRevC.90.061601,PhysRevLett.114.212502,PhysRevC.91.021301} Hamiltonians, 
is here generalized to the treatment of three-cluster dynamics.
We further explore the role of six-body correlations in the description of the low-lying $\alpha$+$n$+$n$ continuum, required to 
accurately evaluate the $^4$He$(2n,\gamma)^6$He radiative capture (one of the mechanism by which stars can overcome the instability of the five- and eight-nucleon systems and create heavier nuclei~\cite{doi:10.1146/annurev.nucl.48.1.175}) and of the $^3$H$(^3$H$,2n)^4$He reaction 
contributing to the neutron yield in inertial confinement fusion experiments~\cite{PhysRevLett.109.025003,PhysRevLett.111.052501}.

\paragraph{Approach.}
In the NCSMC, 
the $A$-nucleon wave function of a system characterized by a {\em core}+$n$+$n$ 
asymptotic in the total angular momentum, parity and isospin channel $J^{\pi}T$ is written as the generalized cluster expansion
\begin{align}
\label{eq:trialwf}
        |\Psi^{J^\pi T}\rangle & =\sum_{\lambda}c^{J^\pi T}_{\lambda}|A\lambda J^{\pi}T\rangle \\ &+ \sum_{\nu} \iint dx \, dy \,   
x^2\, y^2 \, G_{\nu}^{J^\pi T}(x,y) \, \hat {\mathcal A}_\nu\, |\Phi^{J^\pi T}_{\nu x y} \rangle \nonumber \,,  
\end{align}  
where $c^{J^\pi T}_\lambda$ and $G_{\nu}^{J^\pi T}(x,y)$ are, respectively, 
discrete and continuous variational amplitudes to be determined,   
$|A\lambda J^{\pi}T\rangle$ is the square-integrable (antisymmetric) solution for the $\lambda$-th  energy eigenstate of the 
system obtained  working within the $A$-body harmonic oscillator (HO) basis of the 
no-core shell model (NCSM)~\cite{Navratil2000}, 
\begin{align}
\label{eq:basis}
&|\Phi_{\nu xy}^{J^{\pi}T}\rangle 
\!=\!\Big[\!\big (|A-2~ \lambda_c J^{\pi_c}_c T_c\rangle
\;(|n\rangle \;|n\rangle
)^{(s_{nn}T_{nn})}\!\big)^{(ST)}\\
&\times( Y_{\ell_x}(\hat{\eta}_{nn})Y_{\ell_y}(\hat{\eta}_{c,nn}))^{(L)}\!\Big ]^{(J^{\pi}T)}
\frac{\delta(x-\eta_{nn})}{x\eta_{nn}}
\frac{\delta(y-\eta_{c,nn})}{y\eta_{c,nn}}\nonumber
\end{align}
are continuous microscopic-cluster states~\cite{Quaglioni2013} 
describing the organization of the nucleons into 
 an ($A-2$)-nucleon {\it core} 
 and two neutrons $|n\rangle$, and 
$\hat {\mathcal A}_\nu$ is an appropriate intercluster antisymmetrizer 
introduced to preserve the Pauli exclusion principle.
The {\it core} eigenstates 
 are also computed  in the NCSM, employing the same HO frequency $\hbar\Omega$ and consistent
 number of quanta above the lowest energy configuration  $N_{\rm max}$ used for the $A$-nucleon system.
The states of Eq.~\eqref{eq:basis}  are 
labeled by the quantum numbers $\nu = \{A-2\, \lambda_cJ_c^{\pi_c}T_c; s_{nn} \,T_{nn}\, S \,\ell_x \,\ell_y \, L\}$.
Furthermore,
$\vec\eta_{c,nn}$ $=\eta_{c,nn}\hat\eta_{c,nn}$ and $\vec\eta_{nn}=\eta_{nn}\hat\eta_{nn}$ 
are Jacobi relative coordinates 
proportional to the separation 
between the center of mass (c.m.)\ of the {\it core} and that of the 
residual two neutrons, and to the neutrons' relative position, respectively.

Similar to the binary-cluster case~\cite{PhysRevC.87.034326},
upon orthogonalization of expansion~\eqref{eq:trialwf}, 
we obtain the unknown $c^{J^\pi T}_\lambda$ and $G_{\nu}^{J^\pi T}(x,y)$ amplitudes 
by  solving the Schr\"odinger equation
in the model space spanned by the basis states $|A\lambda J^{\pi}T\rangle$ and 
${\mathcal A}_\nu\, |\Phi^{J^\pi T}_{\nu x y} \rangle$.  
However, given the additional relative coordinate, 
in the three-cluster case we first express 
the continuous amplitudes in the orthogonalized expansion (i.e., the relative-motion wave functions) 
in terms of 
the hyperradius 
$\rho = \sqrt{x^2+y^2}$ and 
hyperangle $\alpha = \arctan{\tfrac xy}$ 
and expand them in the hyperangular basis $\phi^{\ell_x\ell_y}_K(\alpha)$ 
analogously to Ref.~\cite{Quaglioni2013}. 
The $^6$He g.s.\ energy and wave function $|\Psi_{\rm g.s.}\rangle$, 
as well as the matrix elements 
of the $\alpha$+$n$+$n$ scattering matrix 
are 
found by 
matching the orthogonalized form of expansion \eqref{eq:trialwf}
with the known asymptotic behavior of the wave function  using  
an extension of the
microscopic $R$-matrix method on Lagrange mesh 
\cite{Descouvemont:2003ys,Descouvemont:2005rc,PhysRevA.65.052710,Hesse2002184,Hesse199837,PhysRevC.87.034326}. 
We obtain convergence of the hyperangular expansion and $R$-matrix method using the same parameters as in Refs.~\cite{Quaglioni2013,Romero-Redondo:2014fya}.
We then analyze the hyperradial components of the $\alpha$+$n$+$n$  relative motion and preferred spatial 
configurations within the g.s.\ of $^6$He. To this end, we
perform a projection of 
$|\Psi_{\rm g.s.}\rangle$ into the orthogonalized cluster 
basis~\eqref{eq:basis}, i.e.,
\begin{align}
	&\sum_{\nu^\prime}\iint dx^\prime dy^\prime x^{\prime\,2} y^{\prime\,2} {\mathcal N}^{-1/2}_{\nu\nu^\prime}(x,y,x^\prime,y^\prime)\langle\Psi_{\rm g.s.}|{\mathcal A}_{\nu}|\Phi^{J^\pi T}_{\nu^\prime x^\prime y^\prime} \rangle\nonumber\\
	&\qquad\qquad\qquad = \frac{1}{\rho^{5/2}} \sum_{K} \tilde u_{\nu K}(\rho) \phi^{\ell_x\ell_y}_K(\alpha)\,,
	\label{projampl}
\end{align}
where ${\mathcal N}_{\nu\nu^\prime}(x,y,x^\prime,y^\prime)$ is the overlap between the antisymmetrized states~\eqref{eq:basis}~\cite{Quaglioni2013}.
Finally, we obtain the matter ($r_m$) and point-proton ($r_{pp}$) root-mean-square (rms) radii by computing the square root of the expectation values on the g.s.\ wave function of the operators
\begin{equation}
r_m^2\equiv\frac{1}{A}\sum_{i=1}^A r_i^2=\frac{1}{A} \rho^2+\frac{A-2}{A} r_{m}^{2(c)},
\label{rm}  
\end{equation}
and
\begin{equation}
r_{pp}^2\equiv \frac{1}{Z}\sum_{i=1}^A  r_i^2\frac{(1+\tau^{(3)}_i)}{2}= r_{pp}^{2(c)}+ R^{2(c)},           
\label{rpp}
\end{equation} 
respectively. Here $Z$ is the total number of protons, $\tau_i^{(3)}$  is the third component of isospin 
and $r_i$ the distance from the $A$-nucleon c.m.\ of the $i$th nucleon, 
$ r_{m}^{2(c)}$ and $r^{2(c)}_{pp}$ are {\em core} operators defined analogously to Eqs.~\eqref{rm} and \eqref{rpp}, respectively, and 
$R^{(c)}=\sqrt{\tfrac{2}{A(A-2)}}\eta_{c,nn}$ 
is the distance  
between the c.m.\ of the {\it core} and that of the whole system.
The expressions on the far right-hand side of Eqs.~\eqref{rm} and \eqref{rpp} are used to compute the matrix elements involving 
the microscopic-cluster portion of the basis and were specifically derived for {\em core}+$n$+$n$ partitions. 
In particular, the formulation of Eq.~\eqref{rpp} is only valid for the present case of isospin $T_c=0$ {\em core}.
A more detailed account of the formalism will follow in a separate publication~\cite{Romero-Redondo2016}.

\paragraph{Results.}
We start from the chiral 
N$^3$LO nucleon-nucleon ($NN$) interaction of Ref.~\cite{N3LO} 
softened via the similarity renormalization group (SRG) method~\cite{PhysRevC.75.061001,
PhysRevC.77.064003,Wegner1994}, 
which minimizes momentum components above a given resolution scale $\Lambda$.
In particular
we work with 
$\Lambda=2.0$~fm$^{-1}$. In nuclei up to mass number $A=6$ this momentum resolution leads
to binding energies close to experiment despite the omission of the three-nucleon (3$N$) components of the SRG-transformed chiral Hamiltonian
~\cite{PhysRevC.83.034301}, and has been shown to induce negligible (less than $1\%$) two- and higher-body corrections of the $^3$H and $^4$He matter radii
computed with bare operators~\cite{Schuster2014}.  
In the interest of showcasing the vast improvement of the present approach over the use of expansions based
exclusively on 
$\alpha$+$n$+$n$ 
microscopic-cluster states, we also perform calculations with the even softer 
$\Lambda=1.5$ fm$^{-1}$ resolution scale adopted in our earlier studies of Refs.~\cite{Quaglioni2013,Romero-Redondo:2014fya}.
Calculations for $\Lambda$ = 1.5 and 2.0 fm$^{-1}$ were carried out using the same $\hbar\Omega=14$ and 20 MeV HO frequencies of Refs.~\cite{Quaglioni2013,Romero-Redondo:2014fya} and \cite{PhysRevLett.114.212502}, respectively.
All results were obtained including only the $J_c^{\pi_c} T_c=0^+ 0$ g.s.\ of the $\alpha$ particle
and the first four, three and two square-integrable eigenstates of the six-nucleon system for the
$J^\pi=0^+,1^{\pm}$ and $2^+$ channels, respectively. 
\begin{table}[b]
\caption{
Computed $^6$He g.s.\ energies in MeV for the 
$\Lambda=1.5$ 
and
$2.0$ fm$^{-1}$ interactions 
as a function of the absolute HO model space size $N_{\rm tot}=N_{0}+N_{\rm max}$, where $N_0$ 
is the number of quanta shared by the nucleons in their lowest configuration. For 
the $^4$He(g.s.)+$n$+$n$ calculation of Ref.~\cite{Quaglioni2013}, $N_0=0$. However, for the the $p$-shell
$^6$He nucleus within the NCSM and NCSMC, $N_0=2$. 
The last two rows show NCSM extrapolated
results, 
and the experimental value, respectively.} 
\begin{ruledtabular}
\begin{tabular}{c c c c c c} 
 & \multicolumn{3}{c}{$\Lambda=1.5$ fm$^{-1}$}&\multicolumn{2}{c}{$\Lambda=2.0$ fm$^{-1}$}\\  
$N_{\rm tot}$  &Ref.~\cite{Quaglioni2013} & NCSM & NCSMC & NCSM &\!NCSMC\\
\hline
6&  -28.91 & -27.71 & -30.02 & -26.44 & -28.64 \\ 
8&  -28.62 &  -28.95 & -29.69& -27.70 & -28.81\\
10& -28.70 & -29.45 & -29.86& -28.37 & -28.97 \\
12&  -28.70 & -29.66 &-29.86& -28.72 & -29.17\\
\hline
$\infty$ &  ---  & -29.84(4)~\cite{Quaglioni2013} & --- & -29.20(11)~\cite{Saaf2014} & --- \\
\hline
Exp. &    \multicolumn{5}{c}{-29.268}  \\
\end{tabular}
\end{ruledtabular}
\label{energy}
\end{table}
\begin{figure*}[t]
\includegraphics[width=5.2cm,clip=,draft=false]{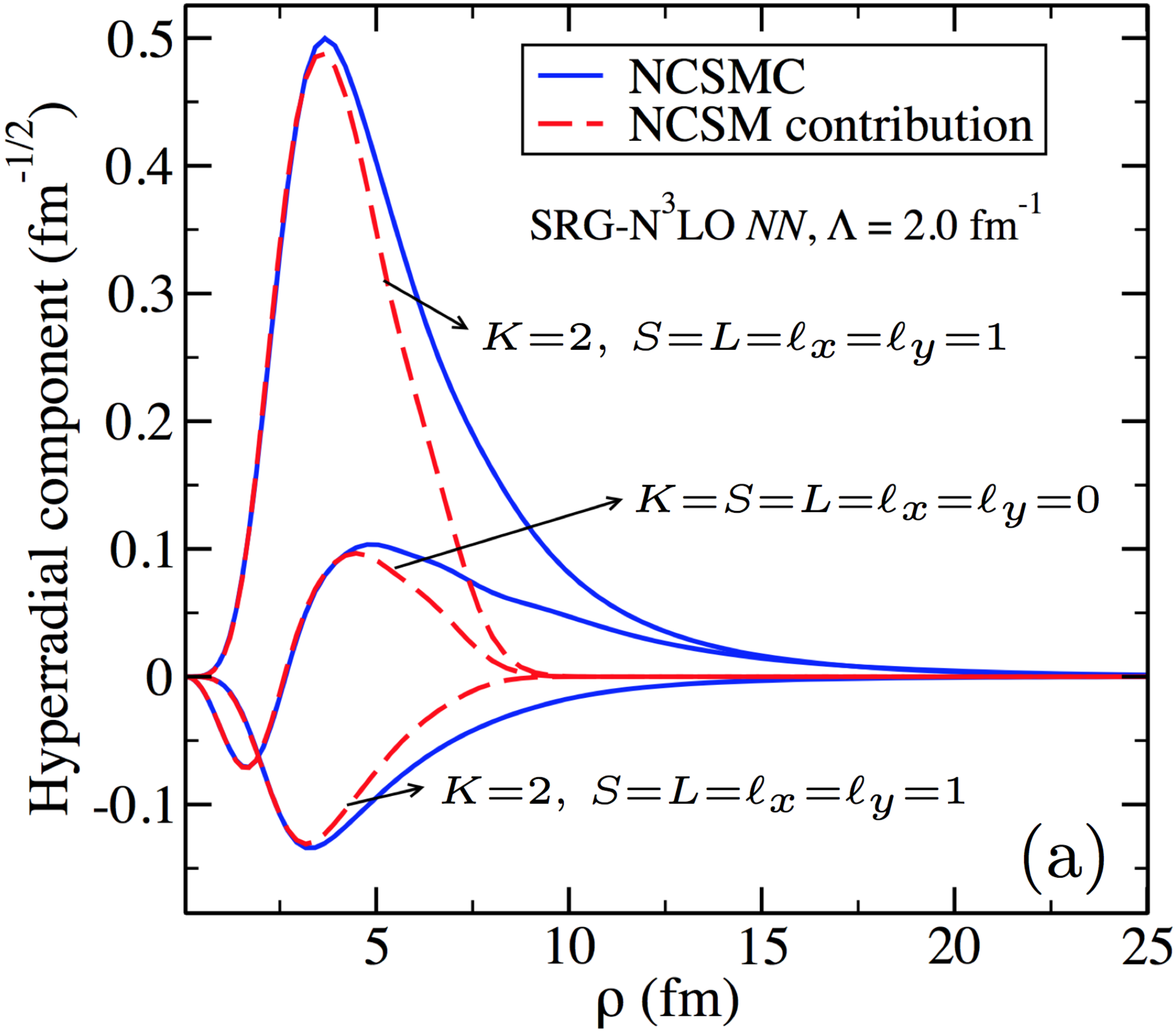}
\qquad
      \includegraphics[width=5.5cm,clip=,draft=false]{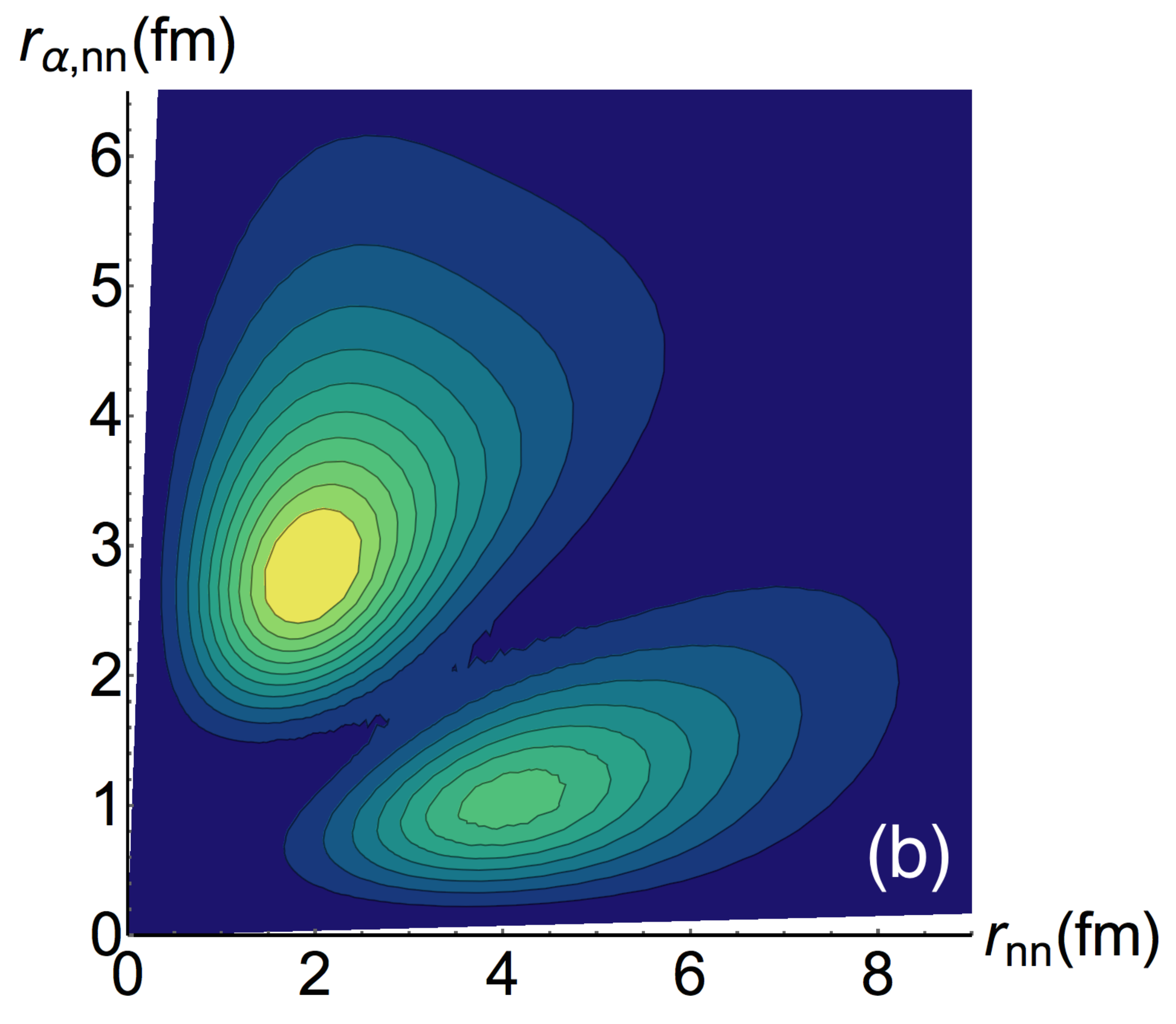} 
      \includegraphics[width=5.5cm,clip=,draft=false]{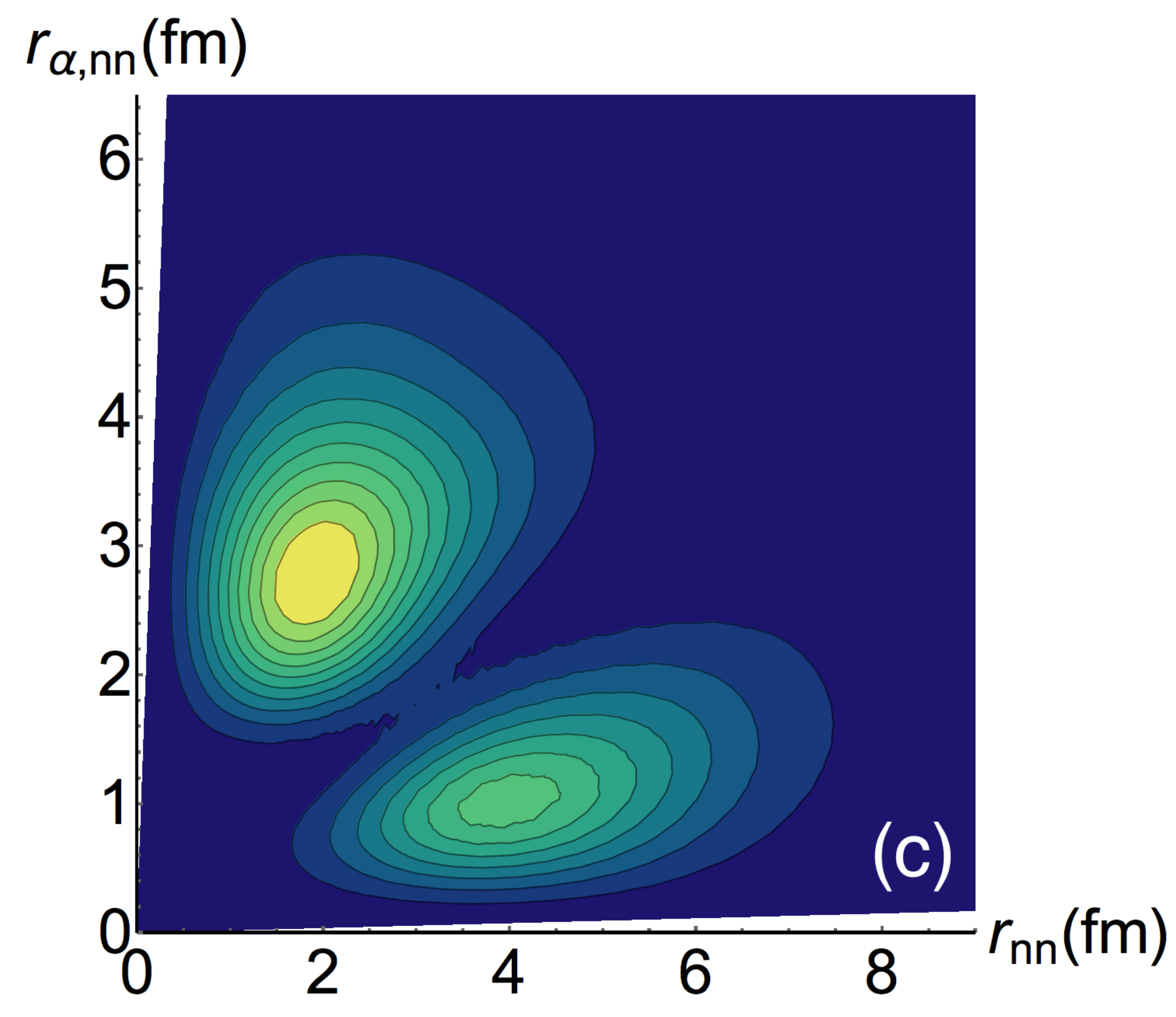} 
\caption{(Color online) Panel (a): Most relevant hyperradial components $\tilde u_{\nu K}(\rho)$ of the $\alpha$+$n$+$n$ relative motion
[see Eq.\eqref{projampl}] within the $^6$He g.s.\ after projection of the $\Lambda=2.0$ fm$^{-1}$ 
full NCSMC wave function in the largest model space (blue solid lines) as well as of its NCSM portion (red dashed lines) 
into the orthogonalized microscopic-cluster basis. 
Panel (b) and (c): Contour plots of the probability distribution 
obtained from the projection of the full NCSMC wave function of panel (a) and its NCSM component, respectively, 
as a function of the relative coordinates $r_{nn}=\sqrt{2}\,\eta_{nn}$ and $r_{\alpha,nn}=\sqrt{3/4}\,\eta_{\alpha,nn}$.
}     
\label{radial_prob}
\end{figure*}

As shown in Table~\ref{energy}, convergence for the $^6$He g.s.\ energy computed within the NCSMC 
is achieved 
within less than 10 keV for $\Lambda=1.5$ fm$^{-1}$, 
and the result is in excellent agreement with the infinite-space extrapolation of the NCSM~\cite{Quaglioni2013}.
This and the good agreement with the accurate extrapolated value of Ref.~\cite{Saaf2014} give us reason to believe that
convergence is achieved in the largest model space considered also for the harder ($\Lambda=2.0$ fm$^{-1}$) interaction. 
In general, the $^4$He(g.s.)+$n$+$n$ degrees of freedom efficiently 
account for the onset of clustering 
already in small model spaces.  
Conversely, 
the square-integrable eigenstates supply 
many-body correlations that 
are not accounted for 
in a microscopic-cluster expansion including only the g.s.\ of $^4$He,
such as the one shown in the first column of the table  
(note that $^6$He is unbound in the analogous calculations for $\Lambda=2.0$ fm$^{-1}$). 
As shown in Fig.~\ref{radial_prob}(a), the $^4$He(g.s.)+$n$+$n$ 
portion of the 
basis serves also the important 
role of providing the correct
asymptotic behavior 
and extended configurations 
of the hyperradial motion 
typical of a Borromean halo 
such as $^6$He. 

The projection over the orthogonalized microscopic-cluster basis of Eq.~\eqref{projampl}
captures 97\% of the original NCSMC solution, confirming the $\alpha$+$n$+$n$ picture of the $^6$He g.s. 
To visualize 
its spatial structure, 
we present in Fig.~\ref{radial_prob}(b) 
the contour plot of the associated probability distribution. 
This displays 
the characteristic dominance of the ``di-neutron" configuration (two neutrons about 2 fm apart orbiting 
 the {\it core} at a distance of about 3 fm) over the ``cigar" picture (two neutrons far from each other 
with the $\alpha$ particle in between)  already seen 
in numerous previous
studies~\cite{Descouvemont:2003ys,
Brida:2010ae,Quaglioni2013,kukulin86,Zhukov1993151,Saaf2014,Nielsen2001373}. 
While these structures 
are already
captured by the square-integrable portion of the basis [see Fig.~\ref{radial_prob}(c)],
they are more spatially extended in the full calculation.
\begin{figure}[b]
\def\svgwidth{310pt}
\begingroup%
  \makeatletter%
  \providecommand\color[2][]{%
    \errmessage{(Inkscape) Color is used for the text in Inkscape, but the package 'color.sty' is not loaded}%
    \renewcommand\color[2][]{}%
  }%
  \providecommand\transparent[1]{%
    \errmessage{(Inkscape) Transparency is used (non-zero) for the text in Inkscape, but the package 'transparent.sty' is not loaded}%
    \renewcommand\transparent[1]{}%
  }%
  \providecommand\rotatebox[2]{#2}%
  \ifx\svgwidth\undefined%
    \setlength{\unitlength}{955.34534534bp}%
    \ifx\svgscale\undefined%
      \relax%
    \else%
      \setlength{\unitlength}{\unitlength * \real{\svgscale}}%
    \fi%
  \else%
    \setlength{\unitlength}{\svgwidth}%
  \fi%
  \global\let\svgwidth\undefined%
  \global\let\svgscale\undefined%
  \makeatother%
  \begin{picture}(1,0.67410178)%
    \put(0,0){\includegraphics[width=\unitlength,page=1]{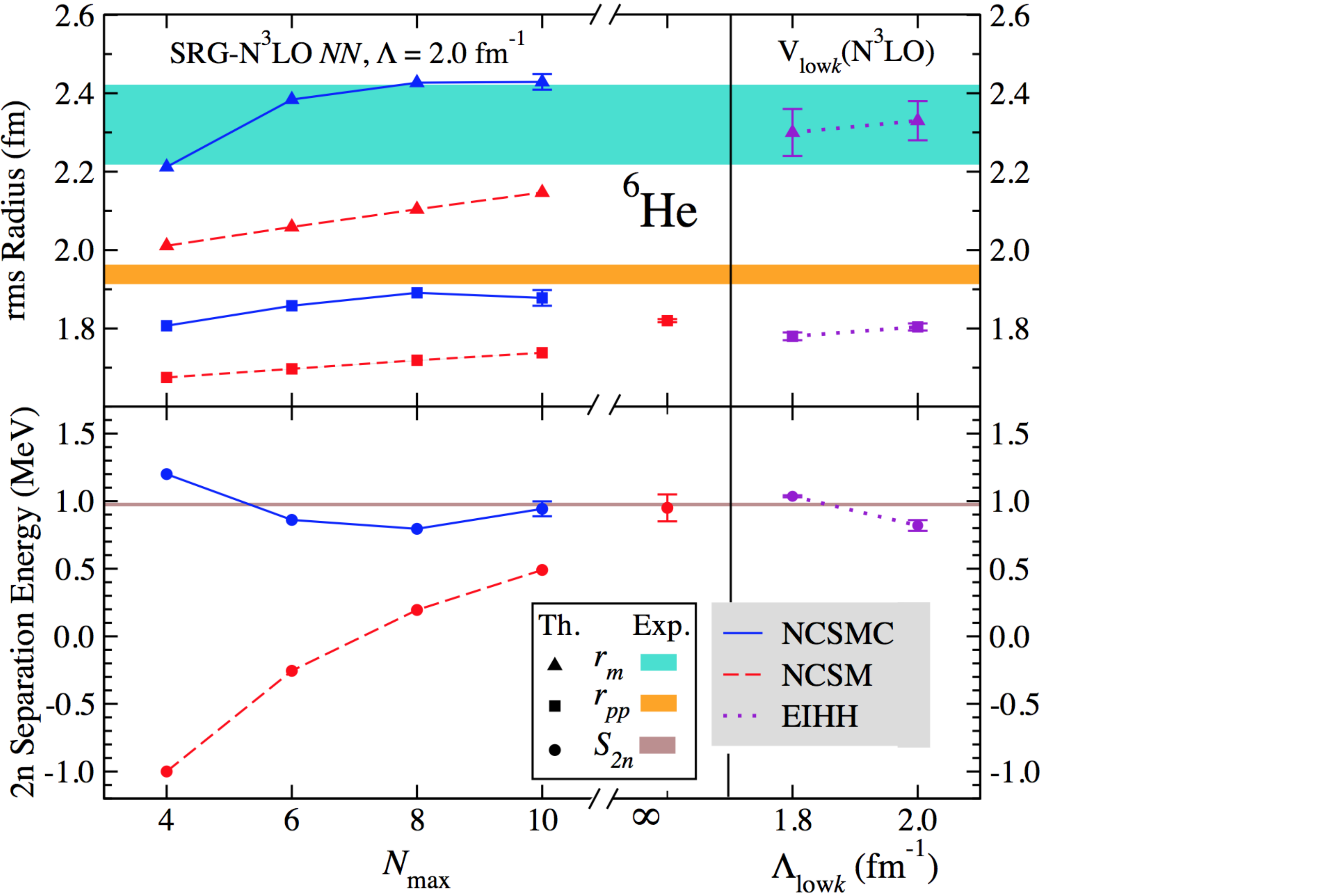}}%
    \put(0.64883445,0.12642301){\color[rgb]{0,0,0}\makebox(0,0)[lb]{\smash{\cite{Bacca:2012up}}}}%
    \put(0.49725275,0.0499729){\color[rgb]{0,0,0}\makebox(0,0)[lb]{\smash{\cite{Saaf2014}}}}%
  \end{picture}%
\endgroup%
\caption{(Color online) NCSMC (blue solid lines) and NCSM (red dashed lines) rms matter 
(triangles) and point-proton (squares) radii, and two-neutron separation energy
(circles), obtained using the SRG-N$^3$LO $NN$ interaction with $\Lambda=2.0$ fm$^{-1}$
as a function of the HO basis size.
Also shown are the 
infinite-basis extrapolations 
from Ref.~\cite{Saaf2014} 
and the 
EIHH results   
from Ref.~\cite{Bacca:2012up} 
at 
the resolution scales $\Lambda_{{\rm low}k}=1.8$, and 2.0 fm$^{-1}$.
The range of experimental values 
are represented by horizontal bands (see text for more details).
}   
\label{fig:conv}
\end{figure}

\begin{table}[b]
\caption{
Summary of the results presented in Fig.~\ref{fig:conv}, with $\Lambda_{{\rm low}k}$ in units of fm$^{-1}$. 
See text 
for further details.  
}
\begin{ruledtabular}
\begin{tabular}{ l l c c c}
           & &$S_{2n}$ (MeV) & $r_m$ (fm) & $r_{pp}$ (fm)\\ 
\hline
NCSMC &($N_{\rm max}=10$)   &  0.94(5)\phantom{1}   & 2.43(2) & 1.88(2) \\
NCSM~\cite{Saaf2014} &($N_{\rm max}=\infty$) &   0.95(10)     &   ---       &\phantom{0}1.820(4)\\
EIHH~\cite{Bacca:2012up} &($\Lambda_{{\rm low}k}=2.0$) &0.82(4)\phantom{1} & 2.33(5) &\phantom{1}1.804(9) \\
\hline
Exp.    &&   0.975\phantom{(1)}   & \phantom{1}2.32(10) & \phantom{10}1.938(23)\\
\end{tabular}
\end{ruledtabular}
\label{summary}
\end{table}
\begin{figure}[b]
      \includegraphics[width=7.5cm,clip=,draft=false]{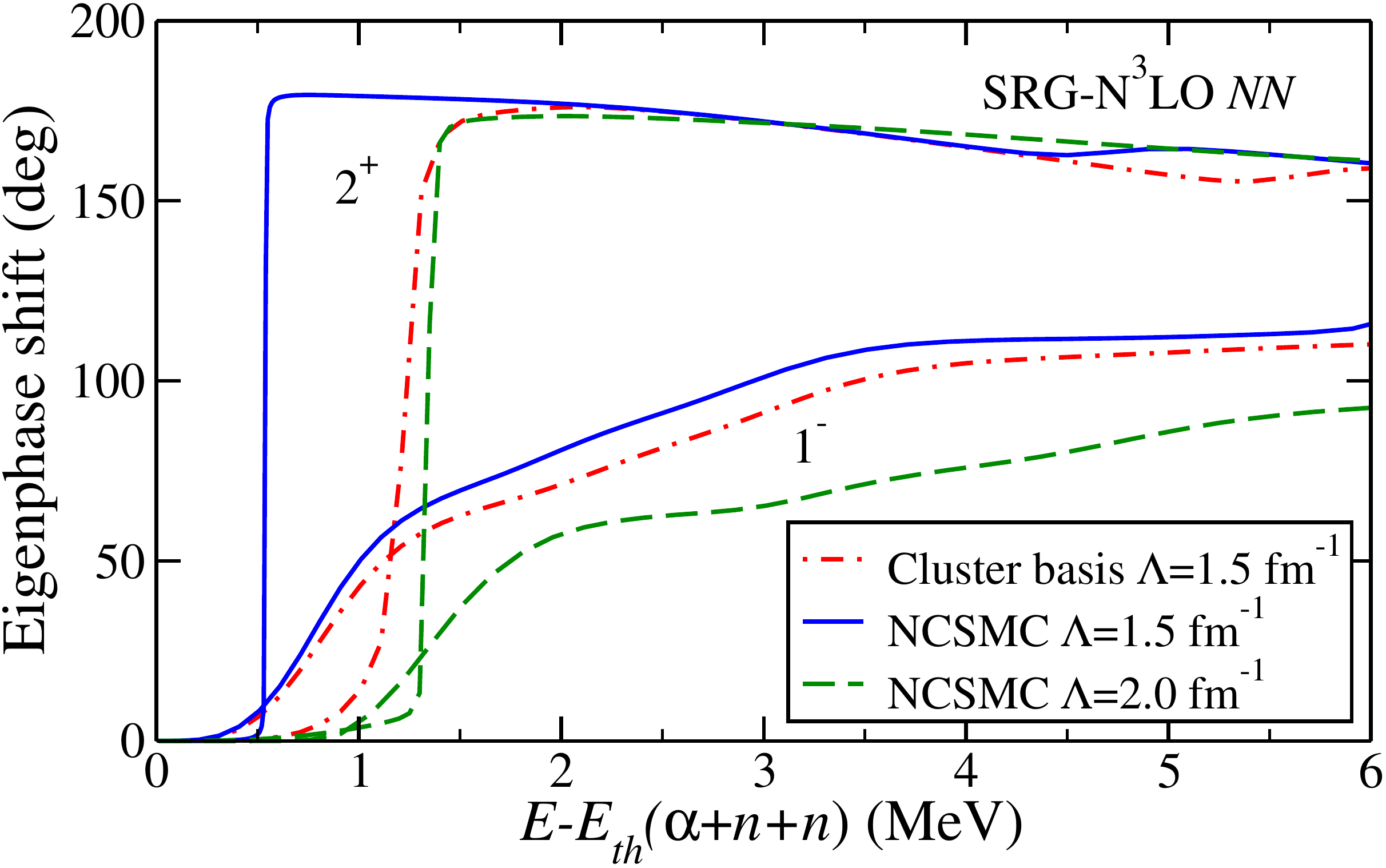}  
\caption{(Color online) Eigenphase shifts for channels $2^+$ and $1^-$ 
as a function of the energy relative to the two-neutron emission threshold 
$E_{th}(\alpha+n+n)$ calculated
with $\Lambda=1.5$~fm$^{-1}$
within the
microscopic cluster basis (red dot-dashed lines) and NCSMC 
(blue solid lines), and $\Lambda=2.0$~fm$^{-1}$ within the NCSMC (green dashed lines).  
 } 
\label{phase_shifts}
\end{figure}

The 
rms matter and point-proton radii obtained from the computed  NCSMC g.s.\ wave functions using the more `realistic' $\Lambda=2.0$ fm$^{-1}$ momentum resolution
are shown together with the corresponding two-neutron separation energy ($S_{2n}$) in Fig.~\ref{fig:conv} and summarized in Table~\ref{summary}.
Also shown as shaded bands are the accurate $S_{2n}$ measurement of Ref.~\cite{PhysRevLett.108.052504}, the range of experimental matter radii spanned by the 
the values and associated error bars of Refs.~\cite{Tanihata:1992wf,Alkhazov:1997zz,kiselev05}, and the bounds for the point-proton radius as evaluated in Ref.~\cite{Bacca:2012up} from the 
charge radius reported in Ref.~\cite{PhysRevLett.93.142501}.
All three observables exhibit 
a considerably weaker dependence on the size of the HO basis 
compared to the results obtained within the NCSM (also shown in the figure as dashed red lines). 
While the NCSMC is not a variational method, given our previous analysis of the binding energy we are confident that the results for the largest model space are reasonably close 
to convergence. An estimate of our uncertainties, based on both the convergence of the two-neutron emission threshold  $E_{th}$($\alpha$+$n$+$n$) and the influence of $^6$He square-integrable states beyond the g.s.\  is reported in Table~\ref{summary} and shown in Fig.~\ref{fig:conv} for the largest model space. There, 
the theoretical 
$S_{2n}$ is closest to its empirical value, 
and the computed $r_m$ 
and $r_{pp}$
radii are, respectively, at the upper end and just below the lower bound of their experimental bands. 
More interestingly, our point-proton radius  is substantially larger than both the extrapolated value of S\"a\"af et al., which ``calls for further investigations"~\cite{Saaf2014}, and the  
effective interaction hyperspherical harmonics (EIHH) result of Bacca et al.~\cite{Bacca:2012up}.  
This latter calculation, based on the $V_{{\rm low}k}$(N$^3$LO) $NN$ interaction, also yields a matter radius smaller than ours though within the experimental bounds. 
The present combination of $S_{2n}$ and $r_{pp}$ values are more in line with the Green's function Monte Carlo results of Ref.~\cite{Pieper2008}, based on $NN$+$3N$ forces constrained to reproduce the properties of light nuclei including $^6$He.

With the present approach we are also able to quantify how the polarization of the $\alpha$ core affects the 
 low-lying continuum of $^6$He, a question that had been left unanswered by our 
previous study~\cite{Romero-Redondo:2014fya}.
At the level of the $^4$He+$n$+$n$ scattering eigenphase shifts obtained for the $\Lambda=1.5$ fm$^{-1}$ momentum resolution, 
the most significant effect is observed for the first $J^\pi=2^+$ resonance, 
which becomes much sharper (with a width of $\Gamma=15$ keV) and is shifted to
lower energies (with the new centroid at 0.536 MeV). This behavior, indicative of a likely influence of the chiral $3N$ force on this state, 
can be seen  in Fig.~\ref{phase_shifts}, which compares results obtained with (NCSMC) and without 
(cluster basis) coupling of $^6$He square-integrable eigenstates. 
The effect in other partial waves is much more moderate. 
In particular, the 1$^-$   
eigenphase shift does not change significantly, excluding core-polarization effects as the possible origin of a low-lying dipole mode. 
\begin{figure}[t]
      \includegraphics[width=8cm,clip=,draft=false]{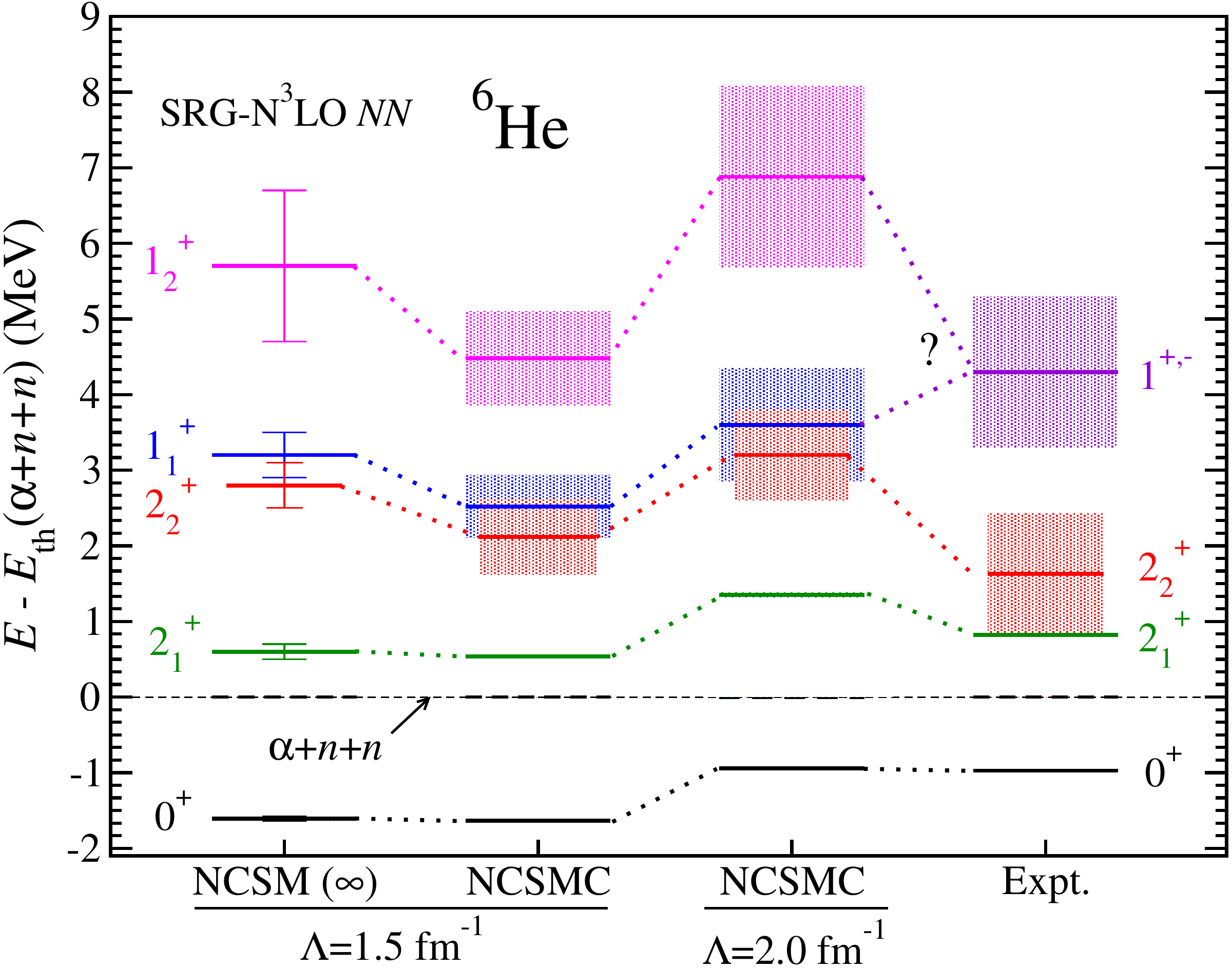}  
\caption{(Color online) Spectrum of $^6$He. Results for
$\Lambda=1.5$ fm$^{-1}$ are shown in the left for both NCSM and NCSMC. 
The third set of  
states corresponds to $\Lambda=2.0$ fm$^{-1}$ within the NCSMC and 
the fourth to the experimental spectrum~\cite{Mougeot:2012aq}.
See text for further details.    
} 
\label{spectrum}
\end{figure}
A summary of the resonance centroids and widths (shown as shaded areas) 
extracted~\cite{Note1,Thompson2009} 
from the computed $\Lambda=1.5$ and $2.0$ fm$^{-1}$  
positive-parity eigenphase shifts
 is presented and compared with experiment~\cite{Mougeot:2012aq} in Fig.~\ref{spectrum}.  
Also shown, for the softer interaction, are 
extrapolated~\cite{Note2} 
energy levels (and their uncertainties) obtained  
within the NCSM by treating the $^6$He excited states as bound states (note that the NCSM does not yield resonance widths). 
Clearly, such an approximation is only justified for the very narrow $2^+_1$ resonance. The two SRG resolution scales produce 
a qualitatively similar picture, with the harder interaction leading to higher-lying and wider resonances (see also Fig.~\ref{phase_shifts}). 

\paragraph{Conclusions.}
We presented a study of $^6$He in which both six-body correlations and $\alpha$+$n$+$n$ clustering are successfully addressed
in the context of the no-core shell model with continuum, 
providing a comprehensive
description of the g.s.\ and low-lying energy continuum of this nucleus. While the inclusion of $3N$ forces 
(currently underway) remains crucial to 
restore the formal unitarity of the adopted SRG transformation of the Hamiltonian and arrive at an accurate description of the
spectrum as a whole, the present results demonstrate that rms matter and point-proton radii compatible 
with experiment can be obtained starting from a soft two-body Hamiltonian that reproduces 
the $^6$He small binding energy.
We conclusively show that a significant portion of the binding energy of the $^6$He g.s.\ 
and the narrow width of the $2^+_1$ resonance are an effect of many-body correlations that, 
in a microscopic-cluster picture, can be interpreted as a consequence of excitations of the $\alpha$
core. This work sets the stage for the {\em ab initio} study of the $^6$He $\beta$-decay half-life 
and $^4$He$(2n,\gamma)^6$He radiative capture, and is a stepping stone in the calculation of the $^3$H$(^3{\rm H},2n)^4$He fusion.

\begin{acknowledgments}
Computing support for this work came from the LLNL institutional Computing Grand Challenge
program. 
Prepared in part by LLNL under Contract DE-AC52-07NA27344. 
This material is based upon work supported by the U.S.\ Department of Energy, Office of Science, Office of Nuclear Physics, 
under Work Proposal No.\ SCW1158, and by the NSERC Grants No.\ 401945-2011 and SAPIN-2016-00033.  
TRIUMF receives federal funding via a contribution agreement with the National Research Council of Canada. 
\end{acknowledgments}

\bibliographystyle{apsrev4-1}

%

\end{document}